\documentclass[journal]{IEEEtran}

\usepackage{dblfloatfix}    

\usepackage{cite}
\usepackage{amsmath,amssymb,amsfonts}
\usepackage{bbm}

\usepackage{graphicx}
\usepackage{epstopdf}
\usepackage{textcomp}
\usepackage{xcolor}
\usepackage{colortbl} 

\usepackage{url} 

\usepackage{bm}
\usepackage{supertabular}
\usepackage{amsthm}
\theoremstyle{definition}

\usepackage[linesnumbered,ruled,vlined]{algorithm2e}
\usepackage{pbox}
\usepackage{float} 
\usepackage{subcaption} 

\usepackage{boldline}
\usepackage{multirow}
\usepackage{array}

\usepackage{enumitem}

\usepackage{pbox}

\usepackage{dblfloatfix}

\usepackage[normalem]{ulem}

\newcommand*\diff{\mathop{}\!\mathrm{d}}

\usepackage{changepage}

\begin{document}

\setlength{\abovedisplayskip}{2pt}
\setlength{\belowdisplayskip}{2pt}

\title{RWP+: A New Random Waypoint Model for High-Speed Mobility
}
%
%
\author{Hussein~A.~Ammar\IEEEauthorrefmark{1},~\IEEEmembership{Student Member,~IEEE}, 
        Raviraj~Adve\IEEEauthorrefmark{1},~\IEEEmembership{Fellow,~IEEE},
        Shahram~Shahbazpanahi\IEEEauthorrefmark{2}\IEEEauthorrefmark{1},~\IEEEmembership{Senior Member,~IEEE},
        Gary~Boudreau\IEEEauthorrefmark{3},~\IEEEmembership{Senior Member,~IEEE},
        and~Kothapalli~Venkata~Srinivas\IEEEauthorrefmark{3},~\IEEEmembership{Member,~IEEE}
        \vspace{-2em}
        \thanks{
                \IEEEauthorrefmark{1}H. A. Ammar and R. Adve are with the Edward S. Rogers Sr. Department of Electrical and Computer Engineering (ECE), University of Toronto, Toronto, ON M5S 3G4, Canada (e-mail: ammarhus@ece.utoronto.ca; rsadve@comm.utoronto.ca).
        }
        \thanks{
                \IEEEauthorrefmark{2}S. Shahbazpanahi is with the Department of Electrical, Computer, and Software Engineering, OntarioTech University, Oshawa, ON L1H 7K4, Canada. He also holds a Status-Only position with the Edward S. Rogers Sr. Department of ECE, University of Toronto.
        }
        \thanks{
                \IEEEauthorrefmark{3}G. Boudreau and K. V. Srinivas are with Ericsson Canada, Ottawa, ON K2K 2V6, Canada.
        }
}


%
%

\maketitle 



\begin{abstract}
In this letter, we emulate real-world statistics for mobility patterns on road systems. We then propose modifications to the assumptions of the random waypoint (RWP) model to better represent high-mobility profiles. We call the model under our new framework as RWP+. Specifically, we  show that the  lengths of the transitions which constitute a trip, are best represented by a lognormal distribution, and that the velocities are best described by a linear combination of normal distributions with different mean values. Compared to the assumptions used in the literature for mobile cellular networks, our modeling provides mobility metrics, such as handoff rates, that better characterize actual emulated trips from the collected statistics.
\end{abstract}

%
\IEEEpeerreviewmaketitle

\section{Introduction}
The random waypoint (RWP) model describes the movement of mobile nodes, e.g., user equipment, in mobile cellular networks. This model was first used by Johnson and Maltz to evaluate dynamic source routing~\cite{johnson1996dynamic}. 
The RWP characterizes many relevant metrics, such as destination and velocity within the set of transitions that represents the whole trip traveled by the user. An assumption frequently used in the RWP model is the mutual independence of the characterized metrics. Specifically, the trip destination and the user speed are chosen independently~\cite{HandoffRateSojourn6477064, Ravi7006787}.

As mentioned in~\cite{8673556}, mobility models suitable for vehicular applications need to be derived. Moreover, high mobility users, e.g., cars and motorcycles, are more likely to make use of road systems. This motivates us to use the mobility pattern on the roads to derive statistics for mobility models such as RWP. Our work fills this gap by providing an accurate characterization of mobility on road systems.

Ours contributions are: 1) Collecting data for road trips; we show  that the transition lengths that constitute  a trip, are best fitted by a lognormal distribution and that the user velocities are best described by a sum of weighted normal distributions with different mean  values. 2) Proposing new framework for the RWP model based on the collected data; we denote the new model with our new framework as RWP+. We show that using RWP+ provides metrics, such as handoff rate, that better describe mobility in a cellular network, compared to the literature~\cite{HandoffRateSojourn6477064}. Supporting reproducible research; we provide data, simulation scenarios and a description of simulations~in~\cite{Ammar_RWP_GT}.

\vspace{-1em}
\section{Mobility Model}\label{section:Model}

\begin{figure*}[t]
        \centering
        \begin{subfigure}[t]{0.242\textwidth}
                \centering
                \includegraphics[width=1\textwidth]{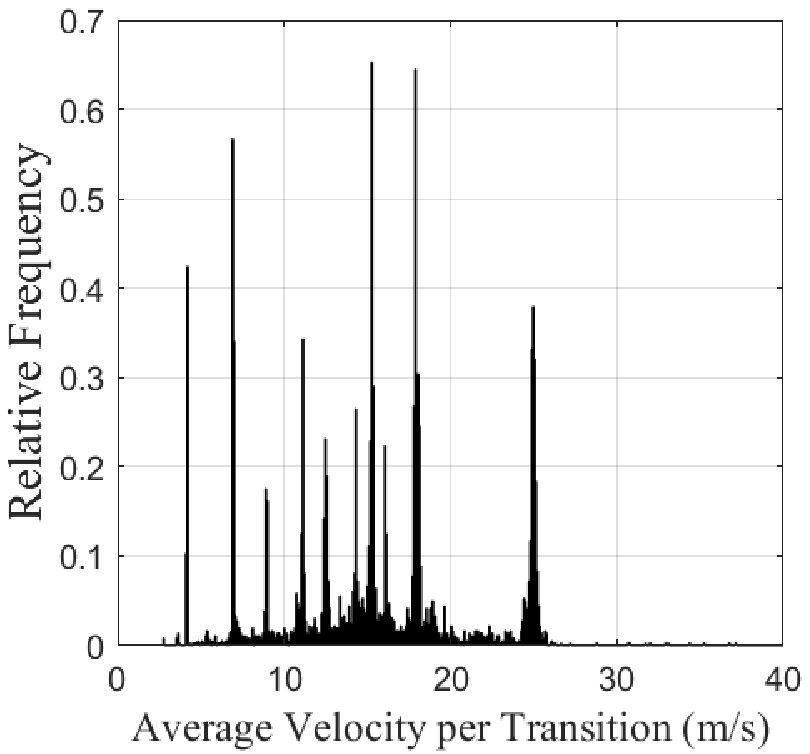}
                \caption{Manhattan area.}
                \label{fig:vPerTransition_manhatten}
        \end{subfigure}
        %
        %
        \begin{subfigure}[t]{0.242\textwidth}
                \centering
                \includegraphics[width=1\textwidth]{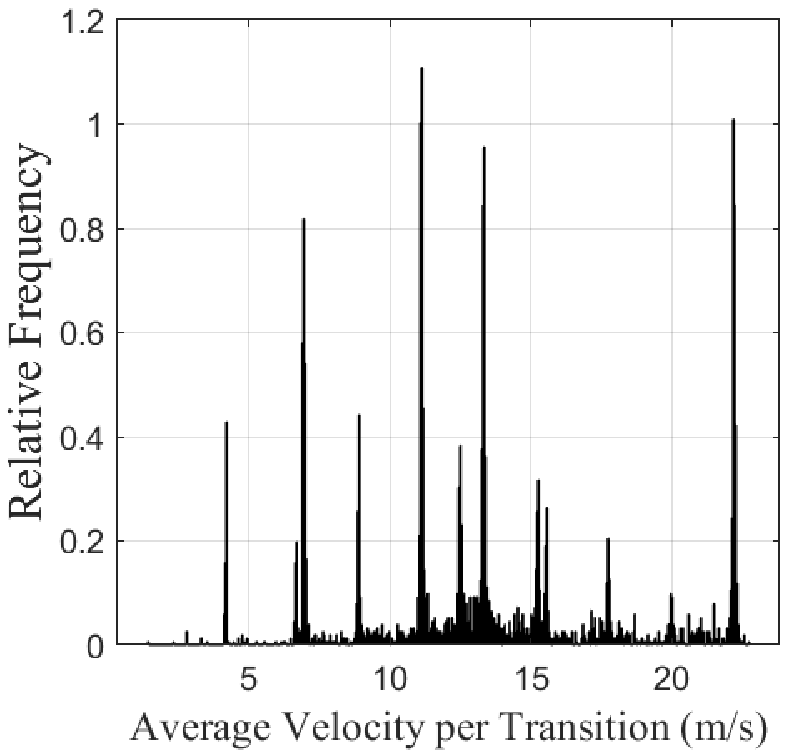}
                \caption{Toronto city area.}
                \label{fig:vPerTransition_toronto}
        \end{subfigure}
        %
        %
        \begin{subfigure}[t]{0.242\textwidth}
                \centering
                \includegraphics[width=1\textwidth]{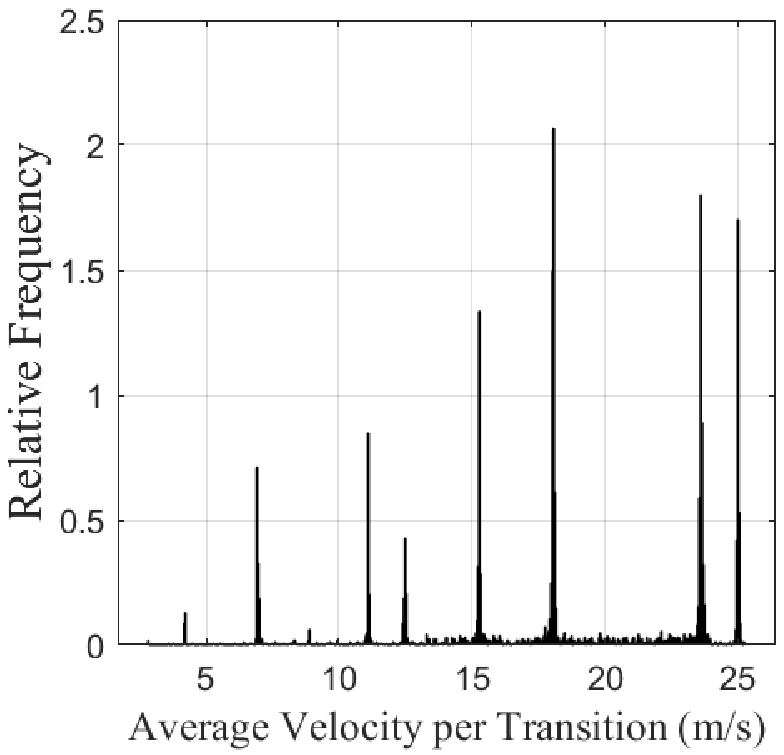}
                \caption{Shanghai city area.}
                \label{fig:vPerTransition_shanghai}
        \end{subfigure}
        %
        %
        \begin{subfigure}[t]{0.242\textwidth}
                \centering
                \includegraphics[width=1\textwidth]{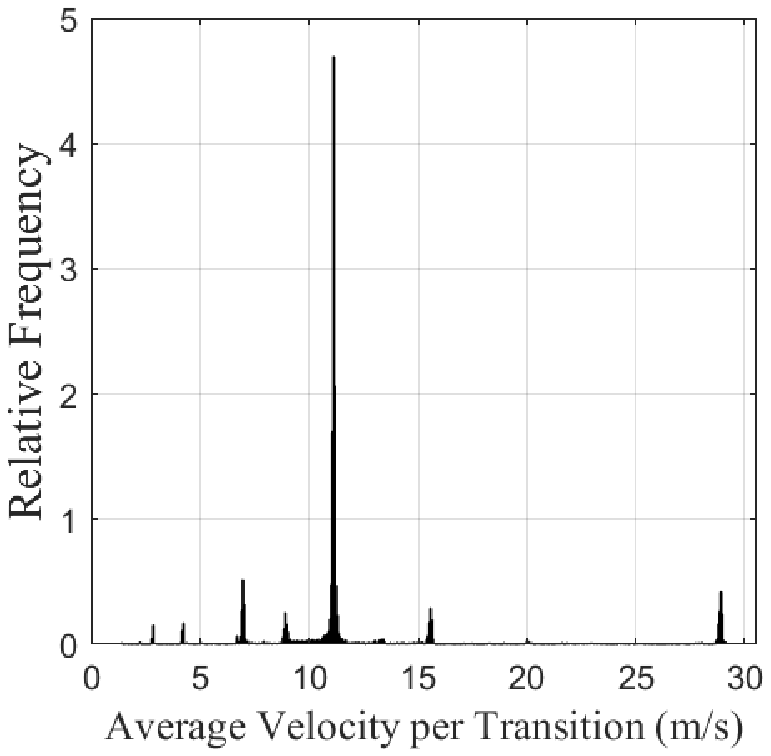}
                \caption{Rome city area.}
                \label{fig:vPerTransition_rome}
        \end{subfigure}
        \vspace{-0.5em}
        \caption{Statistics collected using OSRM: average velocity per transition.}
        \label{fig:stat2}
\end{figure*}
\begin{table*}[t]
        \vspace{-0.5em}
        \scriptsize
        \centering
        \begin{tabular}{|p{0.08\linewidth}|p{0.07\linewidth}|p{0.18\linewidth}|p{0.3\linewidth}|p{0.12\linewidth}|p{0.03\linewidth}|}
                \hline
                \hline
                \multicolumn{1}{|l|}{ \scriptsize \textit{\textbf{Distribution}}}
                &
                \multicolumn{1}{l|}{ \scriptsize \textit{\textbf{Area}}}
                & \multicolumn{1}{l|}{ \scriptsize \textit{\textbf{Parameters}}}
                & \multicolumn{1}{l|}{ \scriptsize \textit{\textbf{95\% Confidence Interval}}}
                &
                \multicolumn{1}{l|}{ \scriptsize \textit{\textbf{RMSE}}}
                &
                \multicolumn{1}{l|}{ \scriptsize \textit{\textbf{Best}}}
                \\
                \hline
                \multirow{4}{1pt}{Exponential} & Manhattan & $\mu$ = 617.87 & $\mu$ = [600.21, 636.32] & 1.094e-4 & 6
                \\
                \cline{2-6}
                & Toronto & $\mu$ = 855.45 & $\mu$ = [827.97, 884.32] & 1.03e-4 & 5
                \\
                \cline{2-6}
                & Shanghai & $\mu$ = 1886.2 & $\mu$ = [1826.02, 1949.42] & 0.314e-4 & 5
                \\
                \cline{2-6}
                & Rome & $\mu$ = 569.23 & $\mu$ = [554.21, 584.87] & 1.13e-4 & 6
                \\
                \hline
                \hline
                \multirow{4}{1pt}{Gamma} & Manhattan & $a$ = 1.26, $b$ = 491.07 & $a$ = [1.21, 1.31], $b$ = [469.31, 513.83] & 0.874e-4 & 2
                \\
                \cline{2-6}
                & Toronto & $a$ = 0.94, $b$ = 910.58 & $a$ = [0.90, 0.98], $b$ = [863.53, 960.20] & 1.042e-4 & 6
                \\
                \cline{2-6}
                & Shanghai & $a$ = 1.29, $b$ = 1464.71 & $a$ = [1.24, 1.34], $b$ = [1392.45, 1540.73] & 0.240e-4 & 1
                \\
                \cline{2-6}
                & Rome & $a$ = 1.03, $b$ = 554.54 & $a$ = [0.99, 1.06], $b$ = [531.28, 578.81] & 1.106e-4 & 5
                \\
                \hline
                \hline
                \multirow{4}{*}{Lognormal} & Manhattan & $\mu$ of log. = 5.98, $\sigma$ of log. = 1.01 & $\mu$ of log. = [5.95, 6.01], $\sigma$ of log. = [0.99, 1.03] & 0.798e-4 & 1
                \\
                \cline{2-6}
                & Toronto & $\mu$ of log. = 6.13, $\sigma$ of log. = 1.13 & $\mu$ of log. = [6.10, 6.17], $\sigma$ of log. = [1.11, 1.16] & 0.459e-4 & 2
                \\
                \cline{2-6}
                & Shanghai & $\mu$ of log. = 7.11, $\sigma$ of log. = 1 & $\mu$ of log. = [7.07, 7.14], $\sigma$ of log. = [0.98, 1.02] & 0.255e-4 & 2
                \\
                \cline{2-6}
                & Rome & $\mu$ of log. = 5.78, $\sigma$ of log. = 1.06 & $\mu$ of log. = [5.76, 5.81], $\sigma$ of log. = [1.04, 1.08] & 0.417e-4 & 1
                \\
                \hline
                \hline
                \multirow{4}{*}{Log-logistic} & Manhattan &$\mu$ of log. = 6.02, $b$ of log. = 0.58  &$\mu$ of log. = [5.99, 6.05], $b$ of log. = [0.57, 0.6] & 0.905e-4 & 3
                \\
                \cline{2-6}
                & Toronto & $\mu$ of log. = 6.12, $b$ of log. = 0.66 & $\mu$ of log. = [6.08, 6.16], $b$ of log. = [0.65, 0.68] & 0.615e-4 & 4
                \\
                \cline{2-6}
                & Shanghai & $\mu$ of log. = 7.15, $b$ of log. = 0.58 & $\mu$ of log. = [7.12, 7.18], $b$ of log. = [0.56, 0.59] & 0.276e-4 & 4
                \\
                \cline{2-6}
                & Rome & $\mu$ of log. = 5.79, $b$ of log. = 0.60 & $\mu$ of log. = [5.76, 5.82], $b$ of log. = [0.59, 0.62] & 0.481e-4 & 2
                \\
                \hline
                \hline
                \multirow{4}{*}{Inverse Gaussian} & Manhattan &$b$ = 617.87, $a$ = 358.14 &$b$ = [594.16, 641.57], $a$ = [343.35, 372.94] & 1.309e-4 & 7
                \\
                \cline{2-6}
                & Toronto & $b$ = 855.45, $a$ = 357.32 & $b$ = [811.88, 899.02], $a$ = [340.69, 373.96] & 0.472e-4 & 3
                \\
                \cline{2-6}
                & Shanghai & $b$ = 1886.20, $a$ = 1117.47 & $b$ = [1806.08, 1966.31], $a$ = [1065.80, 1169.14] & 0.405e-4 & 8
                \\
                \cline{2-6}
                & Rome & $b$ = 569.23, $a$ = 268.31 & $b$ = [546.91, 591.55], $a$ = [258.10, 278.53] & 1.01757e-4 & 4
                \\
                \hline
                \hline
                \multirow{4}{*}{Rayleigh} & Manhattan &$b$ = 609.76 &$b$ = [600.98, 618.8] & 2.887e-4 & 9
                \\
                \cline{2-6}
                & Toronto & $b$ = 978.23 & $b$ = [962.39, 994.60] & 2.477e-4 & 9
                \\
                \cline{2-6}
                & Shanghai & $b$ = 1864.93 & $b$ = [1834.94, 1895.92] & 0.799e-4 & 9
                \\
                \cline{2-6}
                & Rome & $b$ = 685.78 & $b$ = [676.67, 695.14] & 3.308e-4 & 9
                \\
                \hline
                \hline
                \multirow{4}{*}{Nakagami} & Manhattan &$a$ = 0.418, $b$ = 743617.12 &$a$ = [0.405, 0.433], $b$ = [710782.54, 777968.51] & 1.378e-4 & 8
                \\
                \cline{2-6}
                & Toronto & $a$ = 0.31, $b$ = 1913873.07 & $a$ = [0.30, 0.32], $b$ = [1804271.17, 2030132.81] & 1.401e-4 & 8
                \\
                \cline{2-6}
                & Shanghai & $a$ = 0.42, $b$ = 6955909.10 & $a$ = [0.41, 0.44], $b$ = [6614814.73, 7314592.08] & 0.376e-4 & 7
                \\
                \cline{2-6}
                & Rome & $a$ = 0.31, $b$ = 940591.12 & $a$ = [0.30, 0.32], $b$ = [896413.01, 986946.47] & 1.844e-4 & 8
                \\
                \hline
                \hline
                \multirow{4}{*}{Weibull} & Manhattan &$b$ = 643.84, $a$ = 1.11 &$b$ = [626.14, 662.03], $a$ = [1.08, 1.13] & 0.943e-4 & 4
                \\
                \cline{2-6}
                & Toronto & $b$ = 814.08, $a$ = 0.91 & $b$ = [783.53, 845.81], $a$ = [0.89, 0.94] & 1.045e-4 & 7
                \\
                \cline{2-6}
                & Shanghai & $b$ = 1973.69, $a$ = 1.12 & $b$ = [1913.81, 2035.44], $a$ = [1.10, 1.15] & 0.259e-4 & 3
                \\
                \cline{2-6}
                & Rome & $b$ = 551.99, $a$ = 0.94 & $b$ = [535.56, 568.93], $a$ = [0.93, 0.96] & 1.208e-4 & 7
                \\
                \hline
                \hline
                \multirow{4}{1pt}{Birnbaum-Saunders} & Manhattan &$b$ = 368.97, $a$ = 1.14 &$b$ = [358.51, 379.42], $a$ = [1.12, 1.16] & 0.975e-4 & 5
                \\
                \cline{2-6}
                & Toronto & $b$ = 465.68, $a$ = 1.30 & $b$ = [449.50, 481.86], $a$ = [1.27, 1.33] & 0.354e-4 & 1
                \\
                \cline{2-6}
                & Shanghai & $b$ = 1132.72, $a$ = 1.13 & $b$ = [1097.01, 1168.43], $a$ = [1.10, 1.16] & 0.318e-4 & 6
                \\
                \cline{2-6}
                & Rome & $b$ = 321.40, $a$ = 1.24 & $b$ = [312.52, 330.29], $a$ = [1.21, 1.26] & 0.823e-4 & 3
                \\
                \hline
                \hline
        \end{tabular}
        \vspace{-0.5em}
        \caption{Maximum likelihood estimates, legend: $\mu$: mean, $\sigma$: standard deviation, $a$: shape, $b$: scale, $c$: location. For best fitting: (1) best, (9) worst. }
        \label{table:Fitting}   
        \vspace{-2.5em}
\end{table*}

\subsection{Definition}
In the RWP model, a user's route is modeled as a sequence of line segments, called transitions. The $m$-th transition is identified with the starting waypoint ${\bf x}_{m-1}$, the destination waypoint ${\bf x}_{m}$, the velocity $V_{m}$,  
and the pause time at ${\bf x}_{m}$, denoted as $s_{m}$. The model represents the mobility pattern of users in wireless networks through the following set of tuples:
\begin{align}\label{eq:mobModel}
        \mathcal{M} = \left\{ \left( {\bf x}_{m-1}, {\bf x}_{m}, V_{m}, s_{m} \right) \right\}_{m = 1, \dots, M },
\end{align}
where $M$ is the number of transitions.

To construct the transitions of the trip, given ${\bf x}_{m-1}$, we can select ${\bf x}_{m}$ by choosing a direction $\theta_{m}$ and a transition length $L_{m} = \|{\bf x}_{m} - {\bf x}_{m-1}\|$. The angle $\theta_{m}$ is the bearing angle measured clock-wise from the true North. The bearing angle $\theta_{m}$ is often assumed~\cite{HandoffRateSojourn6477064} to be uniformly distributed over the interval $(0, 2\pi]$. Moreover, to maintain the model tractability, it is often assumed that $\{L_{m}, \forall m\}$ are all mutually independent and identically distributed (i.i.d.)~\cite{HandoffRateSojourn6477064, 8673556}.

The authors of~\cite{HandoffRateSojourn6477064} choose $L_{m}$  to follow a Rayleigh distribution with a cumulative density function (CDF) given by $F_L(l)^{(\rm ref.)} = 1 - \exp\left( - \lambda_{\rm wp} \pi l^2 \right),\ l > 0$. However, this assumption is not based on actual statistics, but rather on the intuition that given ${\bf x}_{m-1}$, the waypoint ${\bf x}_{m}$ is selected as the nearest point in a set of points  generated by a Poisson point process (PPP) with density $\lambda_{\rm wp}$. In this paper, we present a better choice for the CDF of $L_m$ that accurately model real world data. To do so, one needs to analyze realistic data, and this is exactly what we do in this letter.
%
%
%
\vspace{-1em}
\subsection{Statistics for Mobility Model}
Most mobility scenarios, especially the ones involving high-speed profiles, occur on road systems. Thus, it is logical to study parameters, such as $L_{m}$ and $V_{m}$, based on real data taken from trips on real-world road systems. To collect such data, we use the open source routing machine (OSRM) platform~\cite{OSRM}. This platform provides the shortest path between any two selected locations on road networks in a chosen area. Such a path is presented as a series of waypoints, which identify transitions. OSRM also provides intermediate points within each transition that identify sub-segments, each of which are associated with a velocity.

Based on these outputs of OSRM, each transition length $L_m$ is calculated as the distance between the consecutive waypoints ${\bf x}_{m-1}$ and ${\bf x}_{m}$. The velocity per transition, $V_m$, is calculated as the average over the velocities in the sub-segments in the $m$-th transition. Mathematically, $\{V_{m}, \forall m\}$ are calculated using the formula $V_{m} = \frac{ L_{m} }{ \sum_{j=1}^{J_m} L_{mj}/V_{mj} }$, where ${J_m}$ is the number of returned sub-segments in transition $m$, and $L_{mj}$ and $V_{mj}$ are the lengths and the velocities within the $j$-th sub-segment of the $m$-th transition. 

We consider four cities with different road layouts. For each city, we use OSRM to produce data corresponding to $200$ trips, each with randomly chosen starting and ending points. In the data, a trip can comprise from a few to about $30$ transitions\label{page:NofTransitionPerTrip}. We then fit candidate CDFs for $\{L_{m}, \forall m\}$ and $\{V_m, \forall m\}$.

For the road system, we choose the following profiles:
\begin{itemize}
\item A typical grid street plan with right angles between the roads found in the area of Manhattan, New York, USA.
\item A typical grid street plan having right angles for main streets with secondary streets that do not necessarily follow the grid system with right angles. The chosen location is the area that includes Toronto in Canada.
\item A less regular street plan that contains a large number of major streets and secondary streets that do not follow a grid system. The chosen location is the area that includes Shanghai in China.
\item A typical street plan that does not follow the grid plan and has irregular city blocks. The chosen location is the city of Rome, Italy.
\end{itemize}
\begin{table*}[t]
        \vspace{-0.5em}
        \scriptsize
        \centering
        \begin{tabular}{|p{0.08\linewidth}|p{0.07\linewidth}|p{0.18\linewidth}|p{0.3\linewidth}|p{0.12\linewidth}|p{0.03\linewidth}|}
                \hline
                \hline
                \multicolumn{1}{|l|}{ \scriptsize \textit{\textbf{Distribution}}}
                &
                \multicolumn{1}{l|}{ \scriptsize \textit{\textbf{Area}}}
                & \multicolumn{1}{l|}{ \scriptsize \textit{\textbf{Parameters}}}
                & \multicolumn{1}{l|}{ \scriptsize \textit{\textbf{95\% Confidence Interval}}}
                &
                \multicolumn{1}{l|}{ \scriptsize \textit{\textbf{RMSE}}}
                &
                \multicolumn{1}{l|}{ \scriptsize \textit{\textbf{Best}}}
                \\
                \hline
                \multirow{2}{1pt}{Exponential} & Manhattan & $\mu$ = 205.47 & $\mu$ = [194.39, 217.54] & 4.411e-04 & 6
                \\
                \cline{2-6}
                & Toronto & $\mu$ = 125.91 & $\mu$ = [119.66, 132.66] & 7.452e-04 & 6
                \\
                \hline
                \hline
                \multirow{2}{1pt}{Gamma} & Manhattan & $a$ = 1.14, $b$ = 180.44 & $a$ = [1.06, 1.22], $b$ = [165.19, 197.09] & 4.248e-04 & 3
                \\
                \cline{2-6}
                & Toronto & $a$ = 1, $b$ = 125.86 & $a$ = [0.94, 1.07], $b$ = [115.91, 136.66] & 7.453e-04 & 7
                \\
                \hline
                \hline
                \multirow{2}{1pt}{Lognormal} & Manhattan & $\mu$ of log. = 4.83, $\sigma$ of log. = 1.04 & $\mu$ of log. = [4.77, 4.88], $\sigma$ of log. = [1, 1.09] & 3.866e-04 & 1
                \\
                \cline{2-6}
                & Toronto & $\mu$ of log. = 4.26, $\sigma$ of log. = 1.13 & $\mu$ of log. = [4.20, 4.32], $\sigma$ of log. = [1.09, 1.18] & 5.615e-04 & 2 
                \\
                \hline
                \hline
                \multirow{2}{100pt}{Log-logistic} & Manhattan &$\mu$ of log. = 4.86, $b$ of log. = 0.6  &$\mu$ of log. = [4.8, 4.91], $b$ of log. = [0.57, 0.63] & 3.984e-04  & 2
                \\
                \cline{2-6}
                & Toronto & $\mu$ of log. = 4.28, $b$ of log. = 0.67 & $\mu$ of log. = [4.22, 4.34], $b$ of log. = [0.64, 0.70] & 6.489e-04 & 3
                \\
                \hline
                \hline
                \multirow{2}{100pt}{Inverse Gaussian} & Manhattan &$b$ = 205.47, $a$ = 105.84 &$b$ = [189.37, 221.58], $a$ = [97.42, 114.26] & 4.806e-04 & 7
                \\
                \cline{2-6}
                & Toronto & $b$ = 125.91, $a$ = 53.74 & $b$ = [115.97, 135.85], $a$ = [49.82, 57.66] & 6.548e-04 & 4
                \\
                \hline
                \hline
                \multirow{2}{1pt}{Rayleigh} & Manhattan &$b$ = 223.58 &$b$ = [217.46, 230.05] & 9.059e-04 & 9
                \\
                \cline{2-6}
                & Toronto & $b$ = 134.3 & $b$ = [130.93, 137.86] & 1.855e-03 & 9
                \\
                \hline
                \hline
                \multirow{2}{1pt}{Nakagami} & Manhattan &$a$ = 0.36, $b$ = 99974.05 &$a$ = [0.34, 0.38], $b$ = [91023.80, 109804.36] & 5.422e-04 & 8
                \\
                \cline{2-6}
                & Toronto & $a$ = 0.34, $b$ = 36073.71 & $a$ = [0.32, 0.36], $b$ = [33027.82, 39400.51] & 9.368e-04 & 8
                \\
                \hline
                \hline
                \multirow{2}{1pt}{Weibull} & Manhattan &$b$ = 207.57, $a$ = 1.02 &$b$ = [195.82, 220.03], $a$ = [0.98, 1.06] & 4.366e-04 & 5
                \\
                \cline{2-6}
                & Toronto & $b$ = 123.59, $a$ = 0.96 & $b$ [116.78, 130.81], $a$ = [0.93, 1.00] & 7.273e-04 & 5
                \\
                \hline
                \hline
                \multirow{2}{1pt}{Birnbaum-Saunders} & Manhattan &$b$ = 118.45, $a$ = 1.2 &$b$ = [111.77, 125.12], $a$ = [1.15, 1.24] & 4.318e-04  & 4
                \\
                \cline{2-6}
                & Toronto & $b$ = 68.29, $a$ = 1.29 & $b$ = [64.60, 71.99], $a$ = [1.24, 1.33] & 4.7e-04 & 1
                \\
                \hline
                \hline
        \end{tabular}
        \caption{Maximum likelihood estimates for data obtained from trips generated inside $1 \times 1~{\rm km}^2$ areas, legend: $\mu$: mean, $\sigma$: standard deviation, $a$: shape, $b$: scale, $c$: location. For best fitting: (1) best, (9) worst.}
        \label{table:Fitting_smallRegion} 
        \vspace{-2em}
\end{table*}

Using the histograms of the transition lengths (not shown here), we obtain, the mean values of the transition lengths, $L_{m}$, as $617.87$, $855.45$, $1886.2$, and $569.23$~meters for Manhattan, Toronto, Shanghai, and Rome respectively. These mean values reflect significantly different road layouts of these cities. Note however that the vast majority (approximately $90\%$) of the transitions are less than 2~km long. Moreover, Fig.~\ref{fig:stat2} presents the histograms of the transition average velocities.
\subsubsection{Fitting CDFs to transition lengths}\label{sec:L}
In Table~\ref{table:Fitting}, we present the root mean square error (RMSE) of fitting candidate CDFs to the values of $L_m$ obtained from OSRM for the aforementioned four cities. In this table we also present the fitted parameters, and ranking of the candidate CDFs. The candidate CDFs are chosen to be continuous, skewed, and only defined for positive numbers. This table shows that the lognormal distribution is the most accurate candidate to model the transition lengths $\{L_m,\forall m\}$, followed by the log-logistic, gamma, Birnbaum-Saunders, Weibull, inverse Gaussian, exponential, Nakagami, then Rayleigh distribution.

To study the validity of the fit in small areas, we consider $200$ small areas chosen randomly inside the cities of Manhattan and Toronto. Each area is of dimension $1 \times 1~{\rm km}^2$. Then, we generate a trip inside each area, i.e., the starting and ending points of the trip are restricted inside this small area. We then use OSRM to obtain the shortest path of each trip as described previously. The results of the fitting are detailed in Table~\ref{table:Fitting_smallRegion}, where we also show the RMSE of the fitting. Again, the results show that the lognormal distribution is among the best to model $L_m$. The mean of logarithm, $\mu_l$, and standard deviation of logarithm, $\sigma_l$, are ($(\mu_l, \sigma_l)^{(\text{Manhattan})} = (4.83, 1.04)$ and $(\mu_l, \sigma_l)^{(\text{Toronto})} = (4.26, 1.13)$). These values are slightly smaller than those obtained in Table~\ref{table:Fitting} for large areas. Moreover, the means of $L_m$ are $205.47$ and $125.91$ meters in the chosen small areas inside Manhattan and Toronto, respectively. As expected, these values are smaller than the ones reported when considering larger areas to generate the trips.

Based on the above results, we use the lognormal distribution to model $L_m$ using the probability density function (PDF)
\begin{align}\label{eq:transitionLengthPDF}
        f_{L_{m}}\left( l \right) = \frac{1}{l \sigma_{l} \sqrt{2\pi}}\exp\left( - \frac{(\log(l) - \mu_{l})^2}{ 2\sigma_{l}^2} \right),\ l > 0
\end{align}
\subsubsection{Fitting CDFs to velocities}\label{sec:V}
As clearly seen in Fig.~\ref{fig:stat2}, the velocity $V_{m}$ is not uniformly distributed, as is widely assumed in the literature, e.g.,~\cite{HandoffRateSojourn6477064, bettstetter2004stochastic}. It seems that high mobility users, such as vehicles, tend to drive at certain ranges of speed more than other ranges. Thus, the assumption of $V_{m}$ following a uniform distribution is not realistic. An analysis of the scatter plot of $V_{m}$ versus $L_{m}$ shows that for a specific value of $L_{m}$, the value of $V_{m}$ can change considerably. Also, when $L_{m}$ changes, $V_{m}$ does not follow a specific pattern. Hence, the correlation between $V_{m}$ and $L_{m}$ is very small. This conclusion is further supported by the fact that the traffic on roads plays a significant role in determining the velocities of sub-segment within the transitions. This means that a large $L_{m}$ can have a small $V_{m}$, if the transition has a high traffic profile. Thus, using correlation between the distributions of the two parameters $V_{m}$ and $L_{m}$ without introducing a complex model for the road traffic is not accurate. Indeed, including correlation will just decrease the mathematical tractability of the mobility model, hence precluding the analysis.

Based on this analysis, we propose to use a convex combination of $D$ normal distributions to model $V_m$. Thus, based on Fig.~\ref{fig:stat2}, we construct the following PDF to model $V_m$:
\begin{align}\label{eq:PDFv}
g_{V_m}\left( v \right) =
\resizebox{0.71\columnwidth}{!}
{$ 
        \sum_{d = 1}^{D} \frac{w_d}{\sum_{d' = 1}^{D} w_{d'}} \frac{1}{\sigma_d\sqrt{2 \pi}} \exp\left( -\frac{\left(v - \mu_d\right)^2} {2 \sigma_d^2} \right)
        $}
\end{align}
where $w_d$ is a weight for each normal distribution which is normalized by $1/{\sum_{d' = 1}^{D} w_{d'}}$. 
The set of means $\{\mu_d : d =\{1, \dots D\}\}$ should be chosen sufficiently above zero to prevent inconsistency in the RWP model, otherwise at some movement steps, nodes may become stuck traveling long distances at low speeds (which is a common problem in classic RWP). This prevents the RWP from reaching a steady state because the average speed of nodes consistently decreases over time~\cite{1208967}. Thus, the set $\{\mu_d : d =\{1, \dots D\}\}$ can be chosen to be uniformly distributed between $[v_{\rm min}, v_{\rm max}]$. \newcommand{\varVMin}{In~\cite[Table~I]{1208967}, typical values of $v_{\rm min}$ and $v_{\rm max}$ and the obtained average velocities are shown, where $v_{\rm min}$ is always set larger than $1$~m/s, while for $v_{\rm min} < 4$~m/s, the average velocity achieved from the simulations tends to be slightly lower than the expected average velocity obtained by using a uniform distribution within the considered interval.}\varVMin\ 
With typically small $\sigma_d$, e.g., $0.25$, we do not need to multiply~\eqref{eq:PDFv} with a step function, because the probability of obtaining a negative velocity is close to zero, in other words, $\int_{-\infty}^{\infty} g_{V_m}\left( v \right) \diff v \simeq \int_{0}^{\infty} g_{V_m}\left( v \right) \diff v = 1$ for large values of $\{\frac{\mu_d}{\sigma_d} : d =\{1, \dots D\}\}$ (e.g., check Table~\ref{table:sim_parameters}).  Finally, from~\eqref{eq:PDFv}, by definition, the mean of $V_m$~is
\begin{align}\label{eq:muV}
        \check{\mu}_V = \frac{\sum_{d = 1}^{D} w_d \mu_d}{\sum_{d' = 1}^{D} w_{d'}} 
\end{align}
This completes the modeling of the PDF of the velocities $V_m$.
\vspace{-1em}
\section{Handoff rate}
We now use our analysis to study the applicability of RWP+ model in measuring an important metric related to mobility, namely the handoff rate in cellular networks. To do so, we model the locations of the base stations (BSs) using a 2D PPP $\Phi$ with density $\lambda$, and as in~\cite{HandoffRateSojourn6477064}, we assume that the users are connected to the nearest BS, hence the cells for the BSs form a Voronoi tessellation. The handoff rate is defined as the expected number of handoffs $N$ during one movement period divided by the expected period time~\cite{HandoffRateSojourn6477064}:
\begin{align}\label{eq:H_OriginalFormula}
        H = \frac{\mathbb{E}[N]}{ \mathbb{E}[T] + \mathbb{E}[S] }
\end{align}
where $T$ and $S$ represents the transition duration and the pause time at the destination waypoint, respectively. We note that $S$ is assumed deterministic in the literature, e.g.,~\cite{HandoffRateSojourn6477064}. For notational simplicity, we drop the index $m$. The transition duration is defined as $T = L / V$, hence $L = V T$ and the Jacobian of the transformation is $J = \diff L/\diff T = V$, Then, given a constant velocity $V = v$, the PDF of~$T$~is
\begin{align}
        &h_{T}(t|V = v) = f_{L}(vt|V = v) |J| 
        \nonumber \\
        &\quad \quad
        = \frac{1}{t \sigma_{l} \sqrt{2\pi}}\exp\left( - \frac{(\log(vt) - \mu_{l})^2}{ 2\sigma_{l}^2} \right),\ l > 0
\end{align}
Thus, the PDF of the transition duration ($t>0$) is 
\begin{align}\label{eq:PDFt}
        &h_{T}(t) =
        \resizebox{0.4\columnwidth}{!}
        {$ 
        \int_{0}^{\infty} h_{T}(t|V = v) g_{V}(v) \diff v
        $} 
        = 
        \resizebox{0.38\columnwidth}{!}
        {$
        \frac{1}{2\pi \sigma_{l} t}
        \sum_{d = 1}^{D} \frac{w_d}{ \sigma_d \sum_{d' = 1}^{D} w_{d'}}
        $}
        \nonumber \\
        & \ \times
        \resizebox{0.85\columnwidth}{!}
        {$ \displaystyle
        \int_{0}^{\infty}
        \exp\left( - \frac{(\log(vt) - \mu_{l})^2}{ 2\sigma_{l}^2} \right)
        \exp\left( -\frac{\left(v - \mu_d\right)^2} {2 \sigma_d^2} \right) \diff v
        $}
\end{align}
Hence, the expected value of $T$ is given~by
\begin{align}\label{eq:exp_T}
        & \bar{\mu}_T \equiv \mathbb{E}[T] =
        \resizebox{0.25\columnwidth}{!}
        {$
        \int_{0}^{\infty} t\ h_{T}(t) \diff T
        $}
        \nonumber \\
        &
        =
        \resizebox{0.94\columnwidth}{!}
        {$ \displaystyle
        \frac{\exp(\mu_{l} + \frac{\sigma_{l}^2}{2})}{\sqrt{2 \pi}}
        \sum_{d = 1}^{D} \frac{w_d}{ \sigma_d \sum_{d' = 1}^{D} w_{d'}}
        \int_{0}^{\infty}
        \frac{ 1 }{ v }
        \exp\left( -\frac{\left(v - \mu_d\right)^2} {2 \sigma_d^2} \right) \diff v
        $}
\end{align}
where the integration in~\eqref{eq:exp_T} can be calculated numerically. The expected number of handoffs can be written as
\begin{align}\label{eq:exp_N}
        \mathbb{E}\left[N\right]
        =
        \sum_{n = 0}^{\infty} n q_N(n)
\end{align}
where $q_N(n)$ is the probability mass function of $N$. Unless $q_N(n)$ is known, \eqref{eq:exp_N} is hard to calculate, however, conditioned on a destination waypoint ${\bf x}_{m}$ and a realization of the Voronoi tessellation, the number of handoffs $N$ is the number of intersections of the segment $[{\bf x}_{m-1}, {\bf x}_{m}]$ and the boundary of the Voronoi tessellation constructed by $\Phi$. Thus, the formula in~\eqref{eq:exp_N} can be solved using the generalized argument for Buffon's needle problem~\cite{schroeder1974uffon}. This problem is cast as a geometric probability problem where a needle of length $d = v \Delta t$ is dropped onto a floor with equally spaced parallel lines, and the aim is to find the probability that the needle crosses a line. We note that $v$ is the velocity, and $\Delta t$ is the considered period time. This is the same definition for the occurrence of a handoff, where the needle represents the transition $[{\bf x}_{m-1}, {\bf x}_{m}]$ and the line is the cell boundary. Thus, we can re-write~\eqref{eq:exp_N}~as~\cite{HandoffRateSojourn6477064}
\begin{align}\label{eq:exp_N_2}
        \mathbb{E}\left[N\right]
        =
        \mathbb{E}[V] |\sin(\theta)| \left(\lim_{|\mathcal{A}| \rightarrow \infty} \frac{|\mathcal{B}_{\mathcal{A}}|}{|\mathcal{A}|} \right) \mathbb{E}[T]
\end{align}
where $|\mathcal{A}|$ is the size of the considered area, $|\mathcal{B}_{\mathcal{A}}|$ is the length of the cell boundaries in $\mathcal{A}$, and $|\sin(\theta)|$ is the absolute value of $\sin(\theta)$. Then, $\lim_{|\mathcal{A}| \rightarrow \infty} \frac{|\mathcal{B}_{\mathcal{A}}|}{|\mathcal{A}|}$ is the average length (length intensity) of the cell boundaries in a unit area~\cite{8673556}. From~\cite{7110473}, $\lim_{|\mathcal{A}| \rightarrow \infty} \frac{|\mathcal{B}_{\mathcal{A}}|}{|\mathcal{A}|} = 2\sqrt{\lambda}$ when the BSs are distributed according to PPP. Based on this, the expected number of handoffs~is
\begin{align}\label{eq:expectedN}
        \mathbb{E}[N] 
        &=
        2\check{\mu}_V \mathbb{E}[|\sin(\theta)|] \sqrt{\lambda} \bar{\mu}_T\, .
\end{align}
When the bearing angle $\theta$ is uniformly distributed over the interval $(0, 2\pi]$, then $\mathbb{E}[|\sin(\theta)|] = 2/\pi$ and we can write the handoff rate~as
\vspace{-0.5em}
\begin{align}\label{eq:H}
        H = \frac{4 \sqrt{\lambda} \check{\mu}_V \bar{\mu}_T}{ \pi \left( \bar{\mu}_T + \mathbb{E}[S] \right) }
\end{align}
where $\check{\mu}_V$ and $\bar{\mu}_T$ are defined in~\eqref{eq:muV} and~\eqref{eq:exp_T}, respectively. Equation~\eqref{eq:H} results in a different handoff rate from that derived in~\cite[equation~(30)]{HandoffRateSojourn6477064} because of the different assumptions made on $L$ and $V$. Our proposed assumptions allow us to better select the input parameters for~\eqref{eq:muV} and~\eqref{eq:exp_T} which, in turn, lead to handoff rates that are close to the actual rates in an actual network deployment. In Section~\ref{sec:results}, we clearly show this contribution by showing that our proposed assumptions lead to handoff rates close to those obtained when using empirical data than the assumptions that are being used in the literature. When the location of the BSs is rotationally invariant, e.g., locations are modeled through a homogeneous PPP, the assumption on the distribution of $\theta$ does not matter because the direction of the movement has no relevance.

\vspace{-0.5em}
\section{Simulation Results}\label{sec:results}
In this section, we simulate the handoff rate. We consider three different analyses, denoted as theoretical, simulation, and empirical. The theoretical (theor.) handoff rate is calculated using~\eqref{eq:H}. The input terms $\check{\mu}_V$ and $\bar{\mu}_T$ are chosen to match with the studied area from the four city profiles. Namely, $\check{\mu}_V$ is calculated using~\eqref{eq:muV}, where the parameters $D$ (the number of normal distributions consisting the PDF of $V_m$), $\{\sigma_d,\forall d\}$, $\{\mu_d^{(\text{[AREA NAME]})}, \forall d\}$ and $\{w_d^{(\text{[AREA NAME]})}, \forall d\}$ are chosen to match the PDFs in Fig.~\ref{fig:stat2} and their values are shown in Table~\ref{table:sim_parameters}. The value of $\bar{\mu}_T$ is calculated using~\eqref{eq:exp_T}, where the values of $\mu_{l}$ and $\sigma_{l}$ for the transition length are obtained from Table~\ref{table:Fitting} for the lognormal distribution.
	
The simulated (sim.) results are obtained using Monte Carlo simulations, as described below, to verify the accuracy of the theoretical results. Simulating the RWP+ model:

\emph{Setup:} We simulate $400$ network realizations with areas of $40\times 40~\text{km}^2$. The network area is chosen large enough to guarantee that the user does not reach the network boundary in any trip. Alternatively, network wrap around can be used to simulate small network areas.

\emph{User movement:} In each network realization, a typical user is generated at the center of the network, then the user makes a trip of $10$ transitions. For each transition we generate a random velocity using~\eqref{eq:PDFv} and a random transition time using~\eqref{eq:PDFt}; we then obtain the transition length using $L_m = V_m T_m$. The bearing angle, $\theta_m$, is chosen as either normally or uniformly distributed, and the user location is updated. To simulate a specific area, Manhattan for example, the input parameters for~\eqref{eq:PDFv} and~\eqref{eq:PDFt} are chosen as those obtained from fitting, as shown in Table~\ref{table:sim_parameters}.

\emph{Handoff:} The number of handoffs can be calculated as discussed previously. It is simply the number of intersections between the transition (a straight line between the start and end of transition) and the Voronoi tessellation constructed by $\Phi$ which represents the locations of the BSs. The handoff rate can then be calculated using~\eqref{eq:H_OriginalFormula}. More details for the simulation with the data collected using OSRM can be found in~\cite{Ammar_RWP_GT}.

\begin{table}[t]
        \vspace{-0.5em}
        \scriptsize
        \centering
        \begin{tabular}{|p{0.45\linewidth}|p{0.45\linewidth}|}
                \hline
                \hline
                \multicolumn{1}{|l|}{ \textit{\textbf{Parameter}}} & \multicolumn{1}{l|}{\textit{\textbf{Value}}}\\
                \hline
                $D$; $\{\mu_d^{(\text{Manhattan})}, d = \{1, \dots, D\}\}$ & 11;\ \  [4.5, 7, 8.9, 11.8, 12.5, 14.5, 15.5, 16.5, 18, 20, 25]~m/s
                \\
                $D$; $\{\mu_d^{(\text{Toronto})}, d = \{1, \dots, D\}\}$
                 & 
                11;\ \ [4.2, 7, 9, 11.2, 12.5, 13.4, 15.3, 15.6, 17.8, 20, 23]~m/s
                \\
                $D$; $\{\mu_d^{(\text{Shanghai})}, d = \{1, \dots, D\}\}$ & 9;\ \ [4, 6.5, 8.5, 11, 12.5, 15, 17.8, 23.5, 25]~m/s
                \\
                $D$; $\{\mu_d^{(\text{Rome})}, d = \{1, \dots, D\}\}$ & 8;\ \ [3, 4.2, 7, 9, 12, 16, 20, 29]~m/s
                \\
                \hline
                $\{w_d^{(\text{Manhattan})}, d = \{1, \dots, D\}\}$ & [6.5, 8.5, 2.5, 5, 4, 6, 10, 6, 10, 1, 7]
                \\
                $\{w_d^{(\text{Toronto})}, d = \{1, \dots, D\}\}$ & [4, 7, 4, 10, 4, 9, 3, 3, 2, 1.5, 9]
                \\
                $\{ w_d^{(\text{Shanghai})}, d = \{1, \dots, D\}\}$ & [1, 5, 0.5, 5, 4, 6, 10, 7, 7]
                \\
                $\{w_d^{(\text{Rome})}, d = \{1, \dots, D\}\}$ & [0.5, 0.5, 1, 1, 10, 1, 0.5, 2]
                \\
                \hline
                $(\mu_l, \sigma_l)^{(\text{Manhattan})}$;
                $(\mu_l, \sigma_l)^{(\text{Toronto})}$;
                &
                (5.98, 1.01);\ \  (6.13, 1.13);
                \\
                $(\mu_l, \sigma_l)^{(\text{Shanghai})}$; $(\mu_l, \sigma_l)^{(\text{Rome})}$ & 
                (7.11, 1);\ \  (5.78, 1.06)
                \\
                \hline
                $\{\sigma_d, \forall d\}$ & 0.25
                \\
                \hline
                \hline
        \end{tabular}
        \vspace{-0.5em}
        \caption{Simulation parameters.}
        \label{table:sim_parameters}   
        \vspace{-2em}
\end{table}

We compare the results of  the following experiments: \textbf{i)} Our proposed model assumptions (RWP+) with emulation for the mobility parameters in Manhattan and Rome. Results are denoted as ``Proposed''. \textbf{ii)} Literature profile used in~\cite{HandoffRateSojourn6477064}: $V_m$ is uniformly distributed between $[v_{\rm min}, v_{\rm max}]$, where the boundaries are chosen based on the minimum and maximum value of $\mu_d$ in Table~\ref{table:sim_parameters}. Furthermore, the same value for the mean of the transition lengths as that obtained from our model is used to allow a fair comparison. More specifically,~\cite{HandoffRateSojourn6477064} selects the destination waypoints $\{{\rm x}_{m}, \forall m\}$ as the nearest point of a PPP with density $\lambda_{\rm wp}$ leading to a mean for transition length that equals $\mathbb{E}[L]^{(\rm ref.)} = 1/(2\sqrt{\lambda_{\rm wp}})$. However, from our model the mean of the transition length is $\mathbb{E}[L] = e^{\mu_l + \sigma_l^2/2}$. By setting the two means to be equal to each other, we obtain a density of waypoint $\lambda_{\rm wp} = 1/(4 e^{2(\mu_l + \sigma_l^2/2)})$. By this, we can construct a fair comparison for our model with that in~\cite{HandoffRateSojourn6477064}. We denote these results as ``Literature profile''. \textbf{iii)}  Empirical results using trips obtained from OSRM inside Manhattan and Rome; BSs locations are still generated as PPP (i.e., $\Phi$). Results are denoted as ``Empirical data''.

\begin{figure}[t]
		\vspace{-0.7em}
        \centering
        \begin{subfigure}{0.46\columnwidth}
                \centering
                \includegraphics[width=1\linewidth]{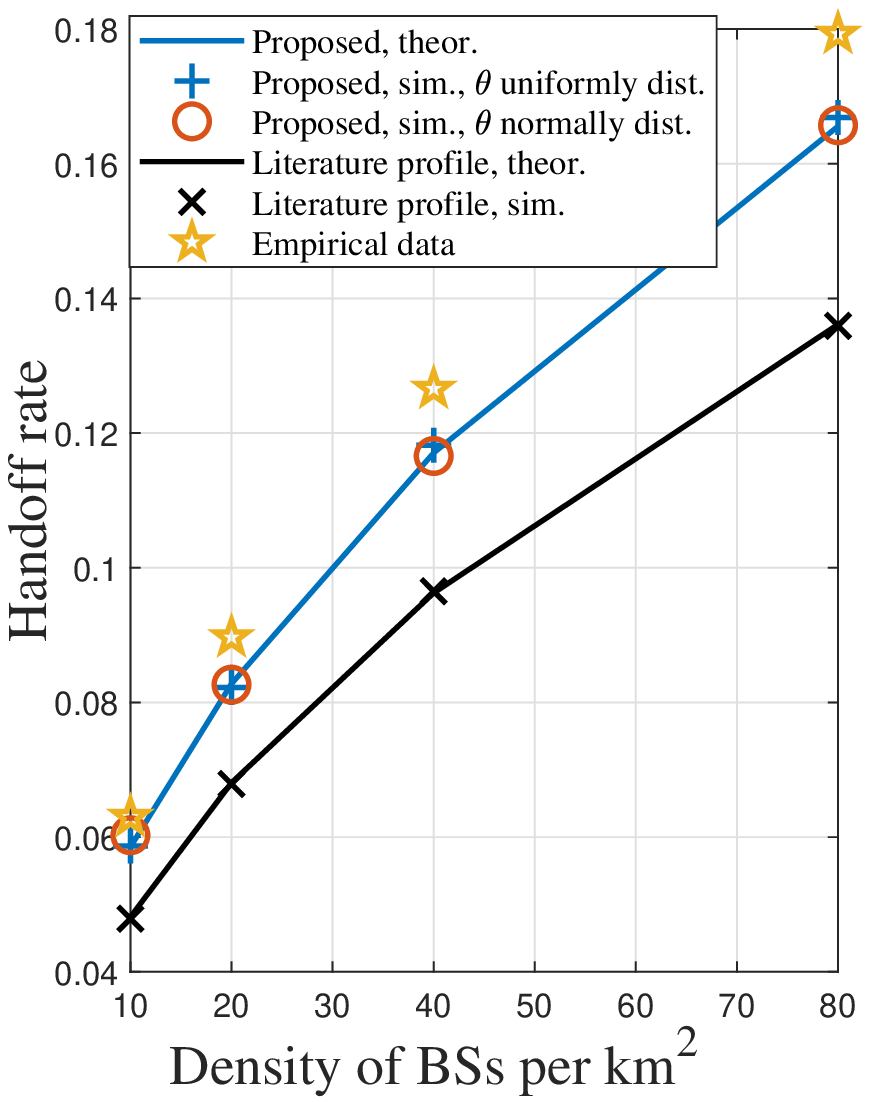}
                \label{eq:HO_rate_manhattan}
        \end{subfigure}
        \begin{subfigure}{0.46\columnwidth}
                \centering
                \includegraphics[width=1\linewidth]{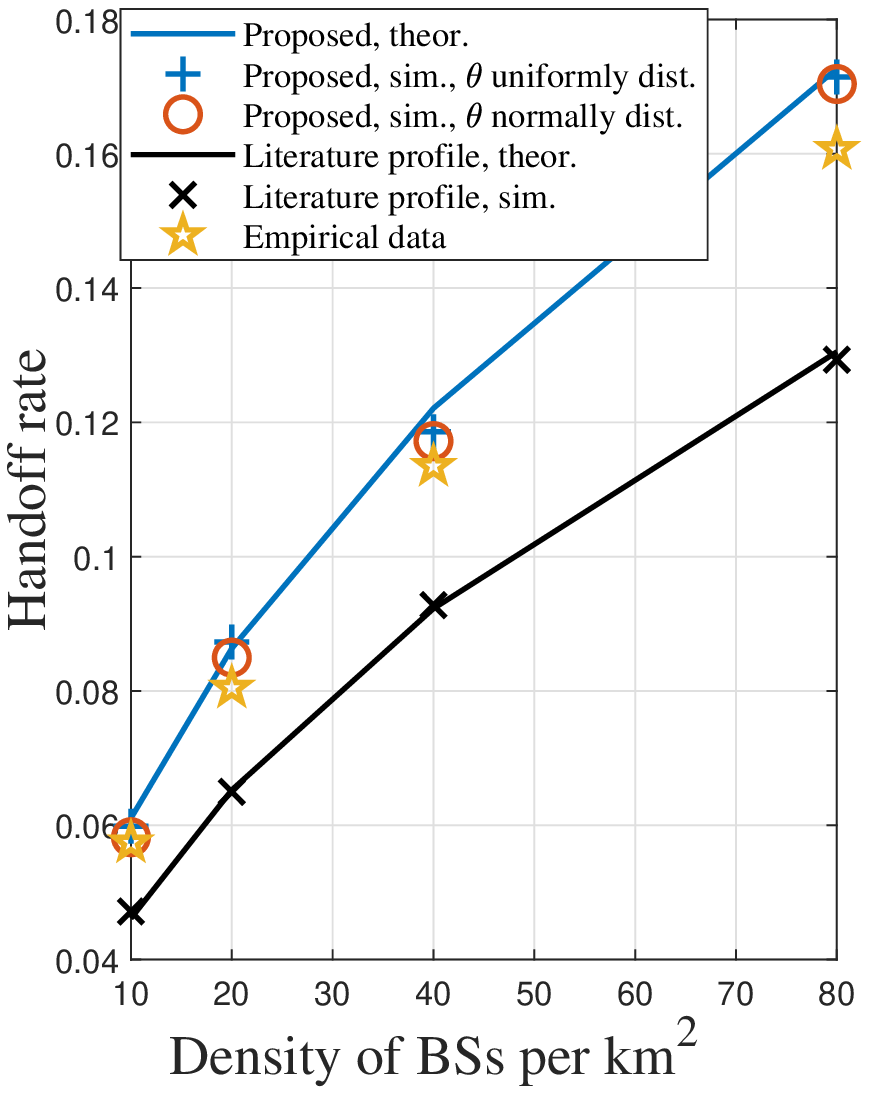}
                \label{eq:HO_rate_shanghai}
        \end{subfigure}
        \vspace{-1.3em}
        \caption{Handoff rate: (left) Manhattan area, $S = 0$; (right) Shanghai city area, $S = 0$~sec.}
        \label{eq:HO_rate}
        \vspace{-1em}
\end{figure}

\begin{figure}[t]
        \centering
        \begin{subfigure}{0.46\columnwidth}
                \centering
                \includegraphics[width=1\linewidth]{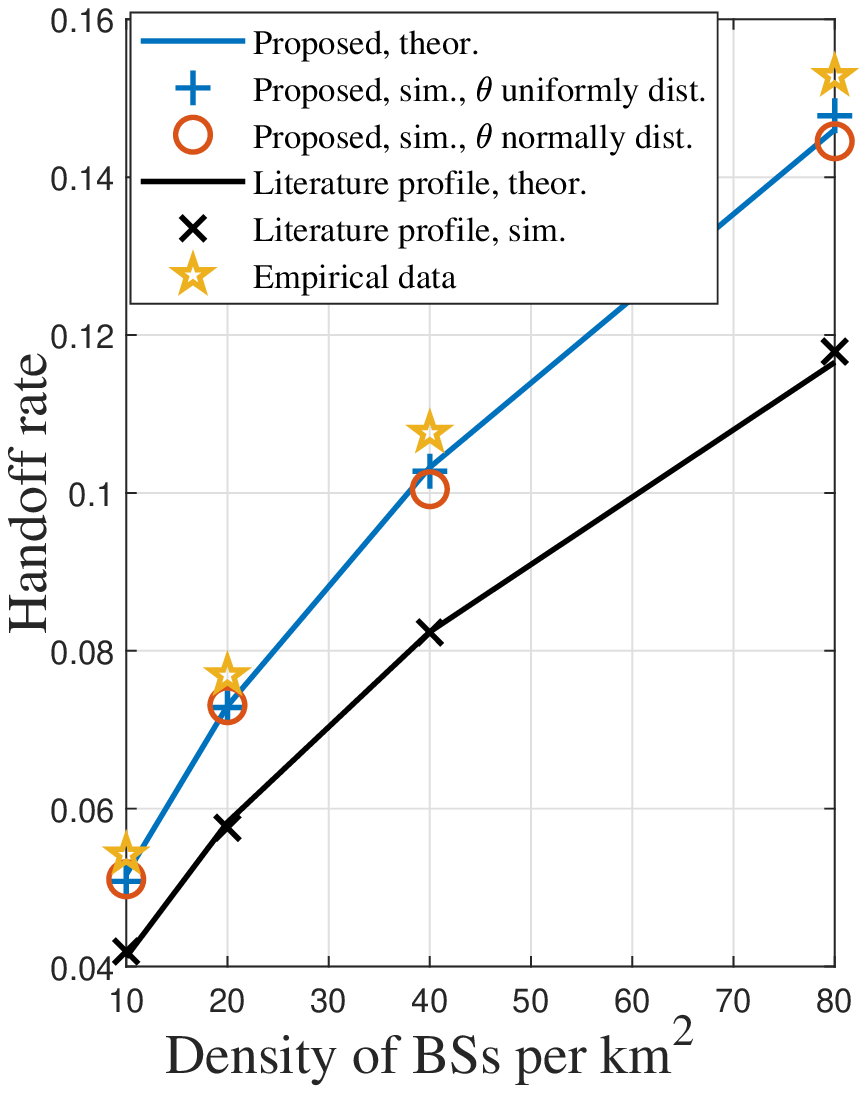}
                \label{eq:HO_rate_toronto}
        \end{subfigure}
        \begin{subfigure}{0.46\columnwidth}
                \centering
                \includegraphics[width=1\linewidth]{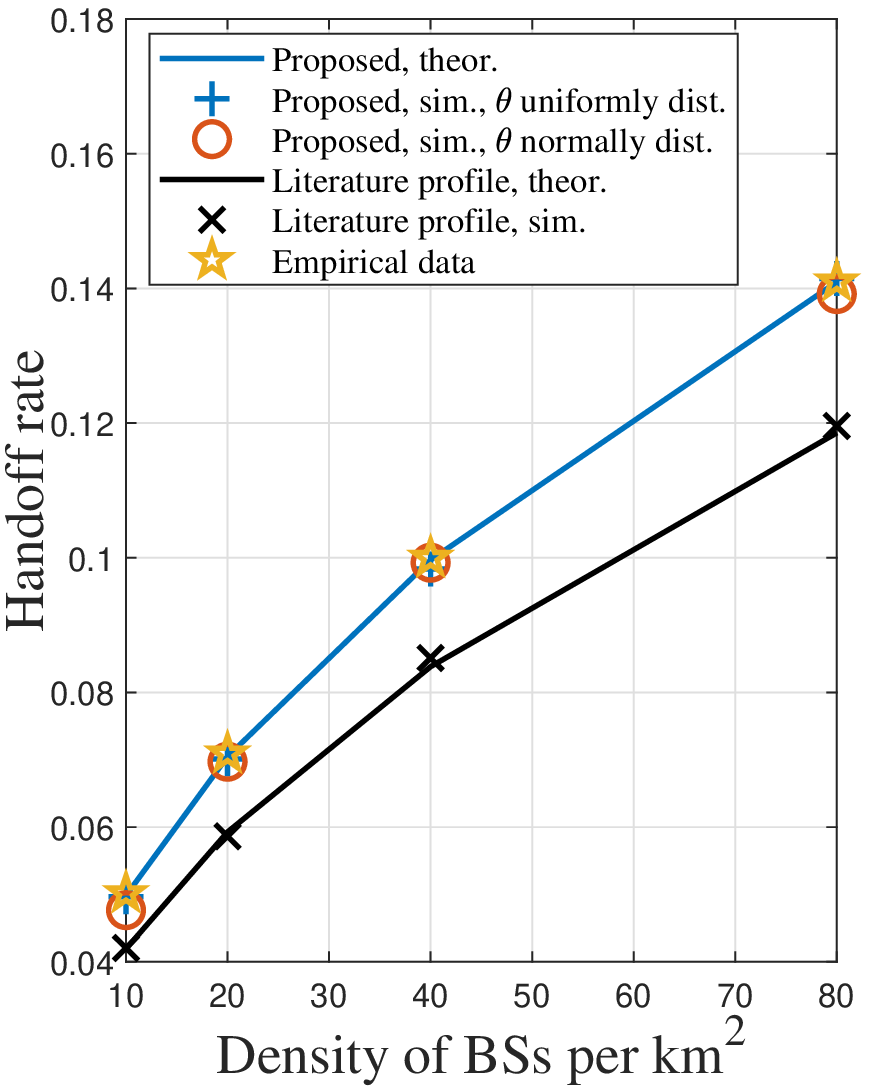}
                \label{eq:HO_rate_rome}
        \end{subfigure}
        \vspace{-1.3em}
        \caption{Handoff rate: (left) Toronto area, $S = 5$;  (right) Rome city area, $S = 5$~sec.}
        \label{eq:HO_rate_2}
        \vspace{-1.5em}
\end{figure}

The results in Figs.~\ref{eq:HO_rate} and~\ref{eq:HO_rate_2} show the accuracy of the theoretical formulas and assumptions compared to Monte Carlo simulation, especially for the assumption of the mutual independence between $V_{m}$ and $L_{m}$. Moreover, by comparing our theoretical results with empirical data, we can clearly see that our new assumptions for the mobility model provide more accurate results compared to those used in the literature, while still maintaining mathematical tractability. This makes the proposed modifications useful for future research into designing handoff policies~\cite{ammar2021user}.


\vspace{-0.5em}
\section{Conclusion}\label{section:conclusion}
We characterized the transition length, velocity and bearing angle for the random waypoint (RWP) mobility model based on real-world statistics. Then, we showed that using our assumptions to derive mobility metrics, such as handoff rate, provides more accurate results for actual mobility scenarios than those obtained from the assumptions used in literature.

\vspace{-0.5em}




%
%
%
%
%
%
\footnotesize
\bibliography{Mob_UC_letter_References}
\bibliographystyle{ieeetr}

\end{document}